# Universes seen by a Chandrasekhar equation in stellar physics

Tom Gehrels*

*Department of Planetary Sciences, University of Arizona*

(Dated: January 12, 2007 )

While we know that quantum, relativity and gravity physics control much of Nature, Subrahmanyan Chandrasekhar derived an equation showing that for the structure, composition, and source of energy of stars. This paper extends its application to universes. A model is derived of these physics indicating that a primordial mass of our universe is finite, at 1.131 79 x $10^{78}$ proton masses. This seems confirmed by two sets of data, from the WMAP spacecraft and other observatories. The model is confirmed more in detail by a determination of the proton radius, at 8.2 (±.2) x $10^{-16}$ m, with a precise theoretical value. This is the equivalent radius for a sphere, while the actual shape of the proton may be ellipsoidal. Together with theories of inflation, the model predicts the existence of a space-time background that is spawning new universes. They all have the same physics and near-critical mass. The multiverse is a hierarchy of increasing numbers of universes. The paper ends with a set of predictions in terms of suggestions for future work.

## I. INTRODUCTION

Subrahmanyan Chandrasekhar (1910-1995) deliberately sought a general equation in terms of cosmological constants as a summary for his theory of the structure, composition and source of energy for stars [1]. He had developed the discipline with detailed expressions such as of Stefan and Boltzmann, relating pressure and temperature at various depths inside the star, but for the total mass of a star he derived

$$M = (hc/G)^{1.5} H^{-2}, \qquad (1)$$

showing that the overall control is by the physics represented by the Planck constant h, the velocity of light c, gravitational constant G, and the mass of the proton H [1, 2]. He noted that, "(t)he essential reason for the success of current theories of stellar structure based on atomic physics can be traced to this fact that $(hc/G)^{1.5} H^{-2}$ is a mass of stellar order". His theory applies also to matter that is degenerate, in the sense of its high density being a function of pressure only, independent of temperature [3]. He also discovered a generalization of Eq. (1),

$$M(\alpha) = (hc/G)^{\alpha} H^{1-2\alpha}, \qquad (2)$$

in which the exponent $\alpha$ identifies the type of object, for instance $\alpha = 1.75$ for our galaxy and 2.00 for our universe, as well as the above $\alpha = 1.50$ for a star. He remarked that this equation might indicate "deeper relations between atomic theory and cosmogony" [2]. He published Eq. (2) in 1937, 1951 and 1989 [1, 2], but he realized that its time had not come for lack of firm data and modeling in cosmology. I took his classes in the 1950s, kept in contact, and remembered the M($\alpha$) concept [4].

---

*E-mail: tgehrels@lpl.arizona.edu

The Wilkinson Microwave Anisotropy Probe (WMAP) and other observing programs have now brought the essential data for using the above equations [5], while theories of inflation have made the time ripe to use them for other universes as well as ours [6,7,8].

This paper begins with a simplification of the equations and a calibration in Sec. II. Section III compares the predictions with observations of objects in our universe. Section IV interfaces M($\alpha$) with inflation theory for their joint application beyond our universe. Section V finds further confirmation of the model, and makes predictions, in terms of proposals for future investigation. Concluding remarks are in Sec. VI.

## II. THE APPLICATION OF M($\alpha$)

Two cosmological prescriptions of M($\alpha$) are specified first. Equation (3) is simplified and calibrated by basing it on the proton and Planck masses. Table 1 will serve as a mainstay for the discussions.

### A. Baryons and original objects only

This paper deals with baryonic masses, *i.e.* consisting of observable matter, which amounts to only 4% of the universe. The other components are 22% dark matter connected with galaxies and 74% in some form of "dark energy" of expanding inter-galactic space [5].

The paper has cosmological meaning only for mass-scaling that happens at the origins of primary masses. In the case of the stars, for instance, the usage of M($\alpha$) will be limited to masses consisting primarily of hydrogen and helium, rather than of the later compositions that have increased abundance of heavier elements in subsequent stars. In fact, those aging stars point towards the universe's demise (Sec. V C).

### B. Usage of the proton and Planck masses

First, a simplification of Eq. (2) is obvious by expressing the masses exclusively in the universal unit of the proton mass, such that H = 1, and

$$M(\alpha) = (hc/G)^{\alpha}, \qquad (3)$$

in proton masses. Even though the H-term has vanished, it is important to remember the presence of the proton in all quantities of mass. The question arises how and why the proton mass is included with h, c, and G. This will become clear after the various aspects of M($\alpha$) come together in Sec. V C. Incidentally, the present treatment and its comparisons with observations indicate that the Planck constant is h, not $\hbar$ = h/2$\pi$; the usage of $\hbar$ may have been initiated by Paul Dirac.

The other essential improvement in the derivation of Eq. (2) is to base it on the Planck mass, which already is a part of cosmology (Sec. IV C). This establishes a powerful physical foundation because it contains h, c, and G together with the involvement of Planck and proton masses. The Planck mass is computed with (hc/G)$^{0.50}$ and, together with the proton mass, they provide a definite calibration for Eq. (3) and its tabulation.

**C. Basic tabulation**

Table 1 uses the above equations, and it is central to the discussions. It presents the data first in proton masses and then in solar masses or kilograms. The class of objects is in the fourth column, and the last column gives the dispersion of the observations in terms of $\alpha$ as will be derived in Sec. III.

**Table 1. Predicted and Observed Values of M($\alpha$) and $\alpha$**

| Exponent $\alpha$ | Predicted proton masses | units shown | Type of Object in M($\alpha$) | Observed $\Delta\alpha$ |
|---|---|---|---|---|
| 2.00 | 1.131 79 (35) x $10^{78}$ | 9.5172 x $10^{20}$ s. m. | Baryonic Universe | 1.998-2.008 |
| 1.75 | 1.981 73 (53) x $10^{68}$ | 1.6664 x $10^{11}$ s. m. | Young Galaxies | 1.72-1.77 |
| 1.50 | 3.469 96 (79) x $10^{58}$ | 29.179 s. m. | Early stars | 1.50-(1.53) |
| 0.50 | 3.261 68 (25) x $10^{19}$ | 5.455 55 (41) x $10^{-8}$ kg | Planck Mass | |
| 0.00 | 1 | 1.672 621 71 (29) x $10^{-27}$ kg | Proton | |

s. m. = solar masses; estimated standard deviations are in the brackets.

Below the center line are the Planck and proton masses, which are obtained from CODATA, combination of laboratory data [9]; h = 6.626 0693 (11) x $10^{-34}$ $m^2$ $s^{-1}$; c = 299 792 458 m $s^{-1}$ (in a vacuum, exact); G = 6.6742 (10) x $10^{-11}$ $m^3$ $kg^{-1}$ $s^{-2}$. The numbers in parentheses are estimated standard deviations; the relative standard uncertainty for G is 1.5 x $10^{-4}$.

### III. COMPARISON WITH OBSERVATIONS

This paper must first establish the capabilities of M($\alpha$) before considering its use outside of our universe, and the way to do that is to check how well it predicts the masses inside. A comparison is therefore made with observations of our universe, galaxies and stars. This Section ends with a critical question and a search for other objects that might be included in the mass scaling.

**A. Our primordial universe**

There are two determinations of baryon density for our universe, and to make the comparison of densities with total mass, one multiplies of course with the volume, but there is a question as to which is the appropriate volume. There is the apparent volume of the presently observable universe, which takes all observational effects into account [10, 11]. However, observational corrections are not applicable to the theoretical and original value of the total baryonic mass in Table 1. The comparison with the observed baryon densities is therefore made without taking the observational effects due to expansion into account, and yet taking the expansion itself into account. That is the volume of a spherical universe having radius 1.373 x $10^{10}$ lightyears, consistent with the expansion-age determination for our universe of 1.373 (+.013, -017) x $10^{10}$ years [5].

The first inferred density was derived theoretically [12], the result ranges between 1.7

and 4.1 x $10^{-28}$ kg m$^{-3}$, yielding between 9.27 x $10^{77}$ and 2.24 x $10^{78}$ proton masses, with α between 1.9979 and 2.0077. A baryon-density observation was made from spacecraft [5] at 4.19 (+.13, -.17) x $10^{-28}$ kg m$^{-3}$, yielding a total mass of 2.300 (+.096,-.126) x $10^{78}$ proton masses, at α = 2.007 89 (+.000 46, -.000 55). The dispersions in α appear to be mostly caused by the uncertainties in the density determinations, the effect of uncertainty in radius 1.373 is less, and this is seen as an indirect confirmation of the expansion age as well as of the comparison method used here. The combined dispersions are in the last column.

The present critical density (for a flat universe, *i.e.* without gravitational collapse or excessive expansion) for baryonic matter is 4.0 (±.4) x $10^{-28}$ kg m$^{-3}$ [Refs. 5 and 13 corrected for $H_0$ of Ref. 5], yielding a total mass of 2.2 (±.2) x $10^{78}$ proton masses, at α = 2.008 (±.001). This mass is 1.9 times the one in Table 1; discussion of the difference is in Secs. V B and C.

### B. Young galaxies

Observations of the 21-cm hydrogen-line for a variety of spiral galaxies show 5 (±4 s.d.) x $10^{10}$ solar masses [14]. One should select the upper limit to allow for dissipation of energy and mass through collisions, while there is also accretion from dwarf galaxies [15]. Carr and Rees derive the upper limit of galaxies near $10^{12}$ solar masses [16]. Observation has also been made of young galaxies at great distance; this is for multiple galaxies occupying a single dark halo, and they total $10^{11}$-$10^{12}$ solar masses [17]. A dispersion in α of only 1.72 – 1.77 is in the Table, representing $10^{10}$ – $10^{12}$ solar masses, even though the observed range for all galaxies, at all ages, has a dispersion of 1.64 – 1.77, representing $10^7$ – $10^{12}$ solar masses. A question whether or not the M(α) model should include the galaxies at such relatively inexact observational status is in Sec. IV C.

### C. Primordial stars

For checking the stars of M(α), physical laws of pressure and temperature within stellar interiors have established that the exponent α is exactly 1.50 [2, 16].

The selection of observations for the present comparison is from stars that have "early" spectral type O, and their values lie near 30 solar masses [18]. They have short lives, and with their supernova explosions, they deliver to the interstellar medium (ISM) atomic nuclei of increased atomic weight, higher than that of hydrogen and helium, a cause of aging of the universe.

Recent reports appear in the literature of more massive stars, but they may be resolved as stellar clusters, or of highly unstable and shedding mass, or have much heavier than hydrogen-and-helium composition, or they consist of accreted masses [19]. Anyway, just to show an extreme of 500 solar masses [20], it is included in the Table as 1.53, but in parentheses to indicate doubt that it fits the criteria of Sec. II A.

Incidentally, "solar mass" in Table 1 merely indicates a unit of 1.9891 x $10^{30}$ kg, rather than numbers of solar-type stars; the early types considered here are fewer and more massive.

### D. Is the mass-scaling real?

The above question is asked in spite of the previous text because there are various publications of scaling for dimensionless numbers and of scaling for various parameters of asteroids, planets, and other objects [21]. There also is misunderstanding that Chandrasekhar participated in Dirac's writing about large-number hypotheses without physical foundation, while the opposite is the case [1].

There is also criticism that the infinite mass in theories of inflation disagrees with the present finite mass, but Sec. III A shows its fits with observations. Furthermore, Ref. (5) already has discussion of the possibility that the universe is finite.

In order to make the $M(\alpha)$ model stand out, an extensive search was made for other primordial objects constrained by Sec. II A. The search was not limited to specific $\alpha$-values because hc/G has the dimension of a mass at all values of $\alpha$.

Stars in open and globular clusters consist mostly of subsequent atomic nuclei, showing spectra later than those of type O.

Clusters of galaxies are included in the data for the second line of Table 1 as subsequent subdivisions. That is how they became grouped, for instance our Milky Way galaxy resides in a supercluster, subdivided into the Virgo Cluster and further into the Local Group. There are however uncertainties in the category of galaxies, and its dispersion is relatively large in Table 1, such that a question of their temporary exclusion from the $M(\alpha)$ model is in Sec. IV C.

At $\alpha = 1.00$, planetesimals of rocks and soil at 1-km radius might appear to be original objects in the solar system, but they do not at all have the type of material considered in Sec. II A, and they came at a late stage in the universe. Stars have the smallest type of primordial objects when they consist of mostly hydrogen and helium and are stable enough for Sec. II A.

No other members were found. Table 1 has all participants of $M(\alpha)$ in our universe.

## IV. OTHER UNIVERSES

Section IV A will confirm Chandrasekhar's confidence in Eq. (2), and introduce the step towards values of $\alpha > 2.00$. Theories of inflation are already involved in such exploration (Sec. B), and Sec. C finds the Planck mass to be a bond between that discipline and $M(\alpha)$. The model presents a series ranked with $\alpha$ for increasing numbers of universes, a hierarchy. Section IV D has notes regarding the proton in order to prepare for Sec. V.

### A. The power of $M(\alpha)$

Equation (3), predicts the mass of our universe close to observations in Sec. III A. Already in the 1930s, Chandrasekhar used $9.5 \times 10^{20}$ solar masses for the universe and, independently of that, $1.1 \times 10^{78}$ for its baryon number. Quantum, relativity, and gravity physics have unified capability we see in Nature especially in the fine-tuning of stellar structure and composition, which, in turn, make atomic nuclei and inorganic and organic evolution possible.

The step beyond $\alpha = 2.00$ takes into account that $M(\alpha)$ is a mass at any value of $\alpha$, as we did in Sec. III D. This type of exploration beyond our universe has no restriction such as the finite velocity of light. Furthermore, inorganic and organic evolution shows that when any option appears, it will be filled [22], and values at $\alpha > 2.00$ are indeed open for $M(\alpha)$, all the way to infinity.

### B. Theories of inflation

Since the late 1970s there has been modeling for a fast expansion of the universe at age $\sim 10^{-33}$ s, $\sim 10^{10}$ Planck times. an "inflation". An intricate procedure is involved with thorough processing towards the beginning of particle formation, removing a variety of uncertainties in the understanding of early stages of our universe. It particularly provides large-scale uniformity and isotropy at that stage because the present 3-K universe displays them [6, 7].

A class of inflation theories has been developed in a well-established discipline, and WMAP observations provide confirmation [5,7]. The models have progressed outside of our universe, such that universes are spawned from a *space-time background* [7,8].

The following Sections use the modeling of such a background and the spawning of universes to base the $M(\alpha)$ model on.

### C. The role and definition of the Planck mass

For a theoretical foundation of inflation, a Planck Era has been discussed [6]. The name of the era refers to a Planck-density phase in which a Planck mass can have most of its components interacting when they would reside within a cubic Planck length, at velocity c (a Planck length in a Planck time). The Planck density is $c^5/hG^2$ and for h, c, and G in Sec. II C it is

$$8.2044\ (18) \times 10^{95}\ \text{kg m}^{-3}. \tag{4}$$

It should be remembered that this is for a Planck mass in a cube with ribs of 1 Planck length. The Planck density may never actually have occurred, but the Planck concepts serve for understanding the root of the theories. The Planck mass connects inflation models with the $M(\alpha)$ model because it too has that theoretical root (Sec. II B). A newly realized definition for the Planck mass appears to be its role in the mass scaling.

For the practical $M(\alpha)$ modeling of consecutive mass-scales, steps of $\Delta\alpha = 0.50$ should presently be taken, rather than of $\Delta\alpha = 0.25$, for two reasons. First, Nature seems to indicate the step of 0.50 at the bottom of Table 1 with the Planck mass at $\Delta\alpha = 0.50$ separation from the proton mass. Second, even though the fit of galaxies to the prediction at $\alpha = 1.75$ in Sec. III B is good relative to its large scaling interval of $\Delta\alpha = 0.25$, galaxies may not be as primary as the stars. There is a question of whether galaxies formed near maximum size or accreted towards that size. It seems prudent to leave the topic for additional observations and interpretations of early galaxies. This causes no difference in the $M(\alpha)$ model, but is merely a temporary caution, and the modeling may later switch to $\Delta\alpha = 0.25$.

### D. The special role of the proton

Andrei Sakharov computed the half-life of the proton at more $10^{50}$ years, but finite [23], and its increasingly difficult observational verification now stands at $10^{35}$ years. Such half-lives seem appropriate for galaxies to have their active lifetime restricted by aging through the transition from hydrogen and helium to heavier elements..

There is something special about the proton. While we expect quantum, relativity and gravity effects with h, c, and G during all the stages of the universe, including the early ones, the proton itself does not appear until age $\sim 10^{37}$ Planck Times, $\sim 10^{-6}$ s. Section V A shows further how the proton is an essential component of the $M(\alpha)$ model, in addition to all masses being expressed in terms of the proton mass.

### V. CONFIRMATION AND FUTURE WORK

Whereas specific predictions by $M(\alpha)$ were verified in Sec. III, additional confirmation is in Secs. V A and B. A set of predictions is in Sec. V C, presented as suggestions for future work.

### A. Determination of the equivalent proton radius

The constant factor F between steps of $\Delta\alpha = 0.50$ in Eq. (3) is seen in the number of proton masses for the Planck mass, which is the same as the numbers of original stars in our universe and of other universes at $\alpha = 2.50$,

$$F = 3.261\ 68\ (25) \times 10^{19}. \tag{5}$$

F also allows the prediction of a size for a universe in the Planck Era, at the Planck density of Eq. 4, at which the Planck mass and the universe compare theoretically.

The ratio of the Planck and universe masses is the third power of F (in Table 1 one sees $\Delta\alpha = 1.50$ difference between $\alpha = 0.50$ and 2.00), but the third root of that is then taken in order to derive the length from the volume ratio, coming back to F. Thus we obtain a size parameter for the universe from F and the Planck length, $3.261\ 68 \times 10^{19} \times 4.051\ 32\ (30) \times 10^{-35} = 1.321\ 41 \times 10^{-15}$ m. That is for a rib of the cube for Eq. (4), while we wish to obtain the radius for the spherical volume of the universe, $R = 8.1974\ (8) \times 10^{-16}$ m. A simpler verification, but without demonstration of the cube in the definition of the Planck density, is to use the mass of the universe (from Table 1, in kg) with the Planck density of Eq. (4), and indeed obtain that radius again. A third derivation is by realizing that for the universe in proton masses $(hc/G)^2$, divided by $c^5/hG^2$, the third-rooted rib value is $h/cp$, where p is the proton mass, and that yields radius R after rib conversion with a precision no longer limited by that of G, but only by that of h and p,

$$R = 8.197\ 3725\ (18) \times 10^{-16}\ \text{m}. \tag{6}$$

The derived numbers appear to be near the average of actually observed proton radii. For a comparison with observations, a straight average of charge-radii obtained by various teams [26] gives $8.2\ (\pm.3) \times 10^{-16}$ m for six observations, of which there is one with $6.4 \times 10^{-16}$ m, while five are between 8.09 and $8.90 \times 10^{-16}$ m. Two more observations yield $8.05\ (\pm.11)$ and $8.62\ (\pm.12) \times 10^{-16}$ m [24].

Equation (6) may be a determination with a new technique, but it is of an *equivalent* proton radius. While the proton has for a long time been considered a fuzzy object, with radii between 6 and $10 \times 10^{-16}$ m, a more interesting interpretation is that it may have a somewhat elliptical shape due to fast quark motion [25, 26]. The word 'equivalent' is then for a hypothetical spherical shape of the proton. One should take into account that Eq. (6) is a theoretical result, which comes from the definition of $(hc/G)^2$ for the mass of the universe, but that as far as its observational verification is concerned, $R = 8.2\ (\pm.2) \times 10^{-16}$ m. The density of the proton, assuming uniformity, follows from Eq. (6) and the mass of the proton in Table 1,

$$\text{Proton density} = 7.249\ 1170\ (50) \times 10^{17}\ \text{kg m}^{-3}. \tag{7}$$

### B. Confirmation and a broad conclusion

The proton result is a confirmation of the Planck parameters in the $M(\alpha)$ model. It confirms the mass of $1.131\ 79 \times 10^{78}$ protons, and it also lies well inside the range of the observations of Ref. 12. On the other hand, the critical density of $2.2\ (\pm.2) \times 10^{78}$ protons (Sec. II A) gives a proton radius of $1.02\ (\pm.03) \times 10^{-15}$ m, which seems unlikely in view of the above observations. A magnification appears by the proton-radius determination of what was a marginal discrepancy in Sec. II A; the observations of Ref. 5 do not have the proton result within their range either. That the theoretical proton radius follows from $\alpha$ = 2.000 comes straightforwardly, but the radius relation is steep so that even $\alpha$ = 2.008 will not give the proper value of R in the Planck domain, and that may indicate a real result (Sec. V C). This paragraph seems to provide a confirmation of the M($\alpha$) model.

The order of time and evolution runs from top to bottom in Table 1. A curious feature is that then, finally near zero-$\alpha$, the listings of Planck and proton masses appear, and they originated not even from cosmological determinations but from h, c, G and H measured in terrestrial laboratories. Their physics must therefore have come to our universe before that, from the outside. We know these to be the quantum, relativity, and gravity physics observed in our present universe. The next two conclusions are then unavoidable, namely that these physics are also the ones of the background space-time from which our universe spawned, and if it happened this way for our universe, it must happen this way for all universes. The broad conclusion follows that the physics and the near-critical mass are the same for every one of the universes in the M($\alpha$) multiverse.

### C. Future work

A practical application of the M($\alpha$) model is to stimulate observations of early galaxies with powerful instrumentation (Sec. III B). The same applies to hydrogen-and-helium stars in order to settle the question of their stability and membership of M($\alpha$) (Sec. III C). A broader quest is for observations of the other universes, remote, indirect, or otherwise; observations of the inter-galactic medium are examples we shall discuss below. As for theoretical applications, the model needs scrutiny and derivation of detailed expressions and predictions. Here are a few examples of practical and theoretical problems that look interesting, but these paragraphs need checking by experts.

A spherically equivalent radius of the proton, with high precision as in Eq. (6), is of interest to particle physicists [25, 26]. It may help determine the shape of the proton through combination with variously observed values of proton size.

The issue of "no beginning" in Refs. (8) and (7) might be re-visited by taking the physics of h, c, G and H into account. The M($\alpha$) model by itself appears to imply that the multiverse is without an ending and without a beginning. There seems to be no scientific way to stop it, nor to have started it all at once, or start it slowly and then increase it. For the same reason, one might surmise that the multiverse is in a steady state. While the potential of $\alpha$ and M($\alpha$) is infinite, it does not preclude the possibility that the multiverse is constant in number. These tentative conclusions need more study, perhaps much later into the future because each era has its restrictions.

Section V A does not actually propose proton size of the universe at Planck density. That volume would be much too large to provide the interaction needed for the observed uniformity of our 3K-universe. For a small enough interaction volume to be considered, even theoretically, its density would have to be $\sim 10^{154}$ kg m$^{-3}$, which is out of the question. Even the Planck density itself would be unrealistic, except perhaps for quantum fluctuations. However, if the space-time background were supplied by the radiation, energy, and matter from aging universes, its density would be much lower. The lead is Sakharov's half-life age limit for the proton (Sec. IV D), which implies a loosening of atomic bonds, increasingly so for the larger nuclei due to their weaker bondage, and thereby yielding an accelerating effect with time. Elementary particles may not decay, or if they do, it should be on an even longer time scale than for the protons, for which Sakharov had already "very large (more than $10^{50}$ years)". The old elementary particles will likely have densities not much larger than the $10^{18}$ kg m$^{-3}$ of Eq. (7), so that may be near the prevailing density of the space-time background. Chandrasekhar's theory of degenerate matter applies at these phases (Sec. I).

One might investigate if there is a closed loop at the space-time background, namely of losses due to the spawning of universes, with matching by gains through accretion from debris of galaxies that are aging due to their transition of hydrogen to heavier elements. Could the rate of spawning new universes be in balance with the rate for demise of old ones? Further study is needed how galaxies come to their end. Does sufficient leakage of radiation, and of energetic, dark, and baryonic matter emerge from the self-gravitation of clusters of galaxies? The former provides radiation pressure, and the others may be enriching the space-time background, but does accretion occur sufficiently? Observations of the inter-galactic medium (IGM) within and outside of clusters of galaxies may be instructive. Here opens another view of our times being ripe, because studies of the intergalactic medium are coming into full swing.

The composition of the space-time background would be uniform because of its supply from leakage of radiation, energy, and matter from various aging galaxies. By the h, c, G, H nature of the aggregates, their end-result in the background is likely to be similar if not the same as before. Modeling of radiation pressure from active galactic nuclei and other bright sources, and perhaps of cosmic rays, are likely to be applicable. There may be useful parallels with the enrichment of the interstellar medium by supernova shockwaves of radiation and matter. There already is exploration of our universe's 4, 22, and 72% compositions.

Delivery to the space-time background seems no problem because of the expansion of the universes. All universes expand eventually through the others. One can see the effect in two dimensions from above a quiet pond on which raindrops fall at various times. The rings expand and travel through each other. This appears to bring effective interaction and many opportunities for energetic accretion within the IUM. Much of this is already in theories of inflation. The M($\alpha$) mass being 1.9 times smaller than the critical value (Sec. V B) may be a real acceleration.

Wherever accretion of the space-time background reaches near-critical mass, it reaches the evolution epoch for scaling at $\alpha$ = 2.00, and the formation of a universe begins. For guidance at understanding of what happens next, the following evolution characteristic is the *spaciousness* we know for our universe, so we see the beginning of expansion controlled among others by the (re-)formation of protons.

An ultimate challenge is the explanation of the mechanisms of mass scaling, evolution, and origins. Why should the formation of a universe begin when the accumulation of baryonic mass for $\alpha$ = 2.00 is reached, at low temperature? Evolution theory deals with this type of question. It may help that simpler questions regarding the maximum mass of galaxies and stars have already been studied [16].

## VI. CONCLUDING REMARKS

Over the years I have followed this topic, new observations always came closer to the predictions of M($\alpha$), and additional corollaries keep coming now. These are good indications of the veracity of the model, as is its beauty [27]. M($\alpha$) is a simple expression bringing a simple model, but it is steeped in physics, Planck, and proton masses. Our universe is a member of an assembly of $3.3 \times 10^{19}$ universes at $\alpha = 2.50$, which is a member of $3.3 \times 10^{19}$ assemblies at $\alpha = 3.00$, and so forth. Not in onion-like shells, for that would imply a center of the multiverse, which there is not, same as there is no center for our universe. The numbering of the alphas is an anthropic view, for we think of our universe at $\alpha = 2.00$, and another society in another universe might do the same. One may visualize the multiverse as a giant tree with many small leaves, seen from anywhere inside. The stems and branches are not continuous, but an image of a dusty day, or of interstellar clouds, would show the space-time background.

A consistent worldview emerges of a finely tuned multiverse with similar universes having the capability of sustaining life, even though no anthropic concept is used. The mass scaling and its atomic theory may be useful in philosophy and evolution as well as in cosmology and particle physics.

## ACKNOWLEDGMENTS

I thank Anthony Aguirre, Neil Gehrels, Piet Hut, Vincent Icke, Carl Koppeschaar, Kevin Moynahan, Masami Ouchi, Kevin Prahar, François Schweizer, Fred Spiers and Erick Weinberg for their valuable comments.

[1] S. Chandrasekhar, Nature **139**, 757 (1937); also in Selected Papers, Vol. I. Stellar Structure and Stellar Atmospheres (Univ. Chicago Press, Chicago, IL,304, 1989).
[2] S. Chandrasekhar, in Astrophysics, a Topical Symposium, edited by J. A. Hynek (McGraw-Hill, New York, N.Y. 598 (1951).
[3] S. Chandrasekhar, Mon. Not. Roy. Astron. Soc. **91**, 456 (1931).
[4] T. Gehrels, On the Glassy Sea, an astronomer's journey (Am. Inst. Phys., New York, NY, 1988), ISBN-0-88318-598-9.
[5] D. N. Spergel et al., Astrophys. J, submitted (2005), astro-ph/0603449.
[6] A. H. Guth, The Inflationary Universe (Addison Wesley, New York, NY, 1997).
[7] A. H. Guth, in Measuring and Modeling the Universe, edited by W. L. Freedman (Cambridge Univ. Pres, Cambridge U.K., 2003), in press, astro-ph/0405546.
[8] A. Aguirre and S. Gratton, Phys. Rev. D **67**, 083515 (2003), gr-qc/0301042.
[9] P. J. Mohr and B. N. Taylor, Rev. Mod. Phys. **77**, 1 (2005).
[10] C. H. Lineweaver and T. M. Davis, Sc. American **292, No. 3**, 36 (2005).
[11] A. Riazuelo, I. P. Uzan, R. Leboucq and J. Weeks, Phys. Rev. D **69**, 103514, Sec. V A (2004).
[12] C. J. Copi, D. N. Schramm and M. S. Turner, Science **267**, 192 (1995).
[13] S. Dodelson, Modern Cosmology (Academic Press, Burlington, MA., 2003)
[14] A. N. Cox, Ed., Allen's Astrophysical Quantities, 4$^{th}$ ed. (Springer Verlag, New York, NY, 579, 2000).
[15] F. Schweizer, F. 2000, Phil. Trans. R. Soc. London A **358,** 2063 (2000).
[16] B. J. Carr and M. J. Rees, M. J., Nature **278,** 605 (1979).
[17] M. Ouchi et al., Astrophys. J. **635,** L117 (2005).
[18] S. W. Stahler, F. Palla and P. T. P. Ho, in Protostars and Planets IV, edited by V. Mannings, A. P. Boss and S. S. Russell (Univ. Ariz. Press, Tucson AZ, 2000), 327.
[19] I. A. Bonnell, S. G. Vine and M. R. Bate, Mon. Not. R. Astron. Soc. (2006), submitted, arXiv:asastro-ph/0401059 v1.
[20] V. Bromm, V., Sky and Tel. **111, No. 5,** 30 (2006).
[21] J. Kleczek, The universe (Reidel Publ., Dordrecht, Netherlands, 1976).
[22] T. Gehrels, Survival through Evolution from Multiverse to Modern Society (BookSurge Publ., North

Charleston, S.C., 2007), in press.
[23] A. D. Sakharov, Zh. Eksp. Teor. Fiz. **49**, 345 (1965); J.Exper.Theor.Phys. **22**, 241 (1966).
[24] S. G. Karshenboim (2000), http://arxiv.org/PS_cache/hep-ph/pdf/0008/0008137.pdf.
[25] D. J. Berkeland, E. A. Hinds and M. G. Boshier, Phys. Rev. Letters **75**, 2470 (1995).
[26] P. F. Schewe and B. Stein, Physics News Update, Phys. Today **242**, Sept. 25 (1995).
[27] S. Chandrasekhar, Truth and Beauty: Aesthetics and Motivations in Science (U. Chicago Press, Chicago, IL, 1987).